%% file: EXO-16-021_temp.tex
\pdfoutput=1

\documentclass[11pt,twoside,a4paper,cmspaper,final,collab]{cms-tdr}
\def\svnVersion{382113}\def\cmsCernNoTag{CERN-EP/2016-230}\def\cmsCernDate{\today}\def\cmsMessage{Published in the Journal of High Energy Physics as \href{http://dx.doi.org/10.1007/JHEP01(2017)076}{\doi{10.1007/JHEP01(2017)076.}}}

\begin{document}\cmsNoteHeader{EXO-16-021}

\hyphenation{had-ron-i-za-tion}
\hyphenation{cal-or-i-me-ter}
\hyphenation{de-vices}
\RCS$Revision: 377291 $
\RCS$HeadURL: svn+ssh://svn.cern.ch/reps/tdr2/papers/EXO-16-021/trunk/EXO-16-021.tex $
\RCS$Id: EXO-16-021.tex 377291 2016-12-14 14:30:58Z pandolf $
\newlength\cmsFigWidth
\ifthenelse{\boolean{cms@external}}{\setlength\cmsFigWidth{0.4\columnwidth}}{\setlength\cmsFigWidth{0.4\textwidth}}

\newcommand{\mzg}{\ensuremath{{M}_{\ell\ell\gamma}}\xspace}
\newcommand{\eeg}{\ensuremath{\Pe^+\Pe^-\gamma}\xspace}
\newcommand{\mmg}{\ensuremath{\mu^+\mu^-\gamma}\xspace}
\providecommand{\PX}{\ensuremath{\mathrm{X}}\xspace}

\cmsNoteHeader{EXO-16-021}
\title{Search for high-mass Z$\gamma$ resonances in  $\Pe^+\Pe^-\gamma$ and $\mu^+\mu^-\gamma$ final states in proton-proton collisions at $\sqrt{s}=8$ and 13\TeV}

\date{\today}

\abstract{
This paper describes the search for a high-mass narrow-width scalar particle
decaying into a Z boson and a photon.
The analysis is performed using proton-proton collision data recorded with
the CMS detector at the LHC
at center-of-mass energies of 8 and 13\TeV, corresponding to
integrated luminosities of 19.7 and
2.7\fbinv, respectively. The
Z bosons are reconstructed from
opposite-sign electron or muon pairs. No statistically significant deviation from the
standard model predictions has been found in the
200--2000\GeV mass range. Upper limits at 95\%
confidence level have been derived on the product
of the scalar particle
production cross section and the
branching fraction of the Z decaying into electrons or
muons,
which range from 280 to 20\unit{fb} for resonance masses between 200
and 2000\GeV.
}

\hypersetup{%
pdfauthor={CMS Collaboration},
pdftitle={Search for high-mass Z-gamma resonances in e+e-gamma and mu+mu-gamma final states in proton-proton collisions at sqrt(s)=8 and 13 TeV},
pdfsubject={CMS},%
pdfkeywords={CMS, physics, Z gamma resonances}}

\maketitle

\section{Introduction}

The ATLAS and CMS experiments have observed~\cite{atlasHiggs, cmsHiggs, long}
a standard model\,(SM) like Higgs boson at
125\GeV~\cite{cms_properties}.
While this
discovery has reaffirmed the SM, it is widely believed that the SM is
a low-energy approximation of a more complex theory~\cite{altarelli}.
An enhancement with respect to the SM in the rate of rare decays of the 125\GeV boson or the discovery
of additional scalar
or pseudoscalar bosons  would  provide evidence that this is the case.
Searches for the rare decay of the 125\GeV Higgs
boson into a \Z boson and a photon have been conducted
 by both ATLAS and CMS~\cite{cms_hzg,Aad:2014fia},
but have insufficient sensitivity to probe the SM Higgs boson hypothesis.

In the context of the wider search for new resonances in the diphoton final
state~\cite{ATLASbump,diphotons,CMSdiphotons2016}, information from the Z$\gamma$ channel provides important complementary information.
For example, an extended SM incorporating a scalar (or pseudoscalar)
decaying to two photons would imply that Z$\gamma$ decays should be observed as well~\cite{Buttazzo:2015txu}.

We present the results for a search for a high-mass scalar, $\PX$, with mass between 200\GeV
and 2\TeV, decaying to $\Z\gamma$.
The analysis is performed by studying proton-proton collisions recorded
with the CMS detector at the CERN LHC. The analyzed data samples correspond
to integrated luminosities of 19.7 and 2.7\fbinv, recorded
at center-of-mass energies of 8 and 13\TeV, respectively.
The search is for localized excesses
in the $\PX \to\Z\gamma$ channel,
with the \Z boson identified by means of its decays
into an electron or a muon pair.
The dominant backgrounds consist of the irreducible contribution from
the continuum $\Z\gamma$  production and the reducible backgrounds from either
final-state radiation in \Z boson decays
or \Z boson production in association with one or more jets (\Z plus jets),
where a jet is misidentified as a photon.
The background is determined directly from data.
Searches for a scalar singlet decaying to $\Z\gamma$ have been
performed at the LHC by ATLAS
at center-of-mass energies of 8~\cite{atlas_hzgHigh}
and 13\TeV~\cite{ATLAS2016}.

\section{The CMS detector}

A detailed description of the CMS detector,
together with the definition of the coordinate system used and the relevant kinematic variables,
can be found elsewhere~\cite{cms}. The central feature of
the CMS apparatus is a superconducting solenoid, 13\unit{m} in length and 6\unit{m} in diameter, which
provides an axial magnetic field of 3.8\unit{T}. Within the field volume there are several particle
detection systems. Charged-particle trajectories are measured by silicon pixel and strip trackers,
covering $0 < \phi < 2\pi$ in azimuth and $\abs{\eta}< 2.5$ in pseudorapidity.
A lead tungstate crystal electromagnetic calorimeter~(ECAL)
is partitioned into a barrel region with $\abs{\eta}< 1.48$ and two endcaps that extend up to $\abs{\eta}= 3$. A
brass and scintillator hadron calorimeter surrounds the ECAL volume and covers the region
$\abs{\eta}< 3$. Iron forward calorimeters with quartz fibers, read out by photomultipliers, extend
the calorimeter coverage up to $\abs{\eta}= 5$. The calorimeters provide measurements of the energy of photons,
electrons, and hadronic jets. Lead and silicon-strip preshower detectors
are  located in front of
the endcap electromagnetic calorimeter. Muons are identified and measured in gas-ionization
detectors embedded in the steel flux-return yoke outside the solenoid. The detector is nearly hermetic,
allowing energy balance measurements in the plane transverse to the beam direction.
A two-tier trigger system selects proton-proton collision events of interest.

\section{Particle reconstruction and event selection}

The selected events are required to pass a dielectron trigger, which has transverse momentum,
$\pt$, thresholds of 17 and 12\GeV, respectively, on the two electrons,
or a dimuon trigger, with thresholds of 17 and 8\GeV on the two muons.
The analysis of the 13\TeV
data also makes use of trigger paths that require the
presence of only one muon, with a transverse momentum threshold of 20\GeV. The trigger
efficiencies for events containing two leptons satisfying the
subsequent event selection requirements
are  measured to be between 90\% and 98\% for the $\eeg$ channel
depending on the electron
transverse momenta, and about  91\% for the $\mmg$ channel.
These
efficiencies are determined with a data sample
enriched in \Z boson events.

Events with two opposite-sign, same-flavor leptons
(electrons or muons) and a photon are selected.
All particles are required to be isolated, and the lepton with the highest
{$\pt$} is required to satisfy $\pt> 20\,(25)$\GeV
in the analysis of 8\,(13)\TeV data,
while the second-leading lepton  must have $\pt>$ 10 (20)\GeV.
The photon is required to satisfy $\pt > 40$\GeV.
The electrons and photon must have $\abs{\eta} < 2.5$,
while the muons must have $\abs{\eta} < 2.4$.
Photons and electrons in the ECAL barrel-endcap
transition region $1.44 < \abs{\eta} < 1.57$ of the electromagnetic
calorimeter are excluded.
More details on reconstruction of photons, electrons, and muons can be
found in Refs.~\cite{photons, CMS:2013ele, muons}.

Events are required to have at least one  vertex~\cite{Veszpremi:2014hpa},
with the  reconstructed longitudinal  position  within 24\unit{cm} of the geometric
center of the detector and the  transverse position within 2\unit{cm} of the beam interaction region. There are
multiple reconstructed  vertices associated with  additional interactions (pileup), and the vertex with the highest  sum of the ${\pt^2}$
of its associated tracks is chosen as the primary vertex. The
leptons  are required to originate from the same primary vertex by requiring, for each track,
that its transverse
impact parameter with respect to the primary vertex is smaller than 2\unit{mm}
and that its longitudinal impact parameter is smaller than
2\,(5)\unit{mm} for electrons~(muons).

The observables used in the photon selection are as follows: isolation
variables based on a particle-flow (PF) algorithm~\cite{pflow,pflow2},
kinematic variables
corresponding to the location and energy of the photon, shower shape
variables that provide information on the
size and shape of the energy deposition in the ECAL, and a
variable taking into account the energy deposited by  pileup
interactions, calculated with the {\FASTJET}
package~\cite{Cacciari:2007fd}.
Identification and isolation requirements in the analysis of the 8\TeV data
are enforced through
the use of a multivariate discriminant, whereas simple, cut-based selection
is used
in the analysis of 13\TeV data.
The search conducted in 8\TeV data targets a lower mass range,
so the photon identification criteria with the most efficient
rejection of the jet-induced background were chosen.

Photon candidates are rejected if a cluster of hits in the tracker pixel
detector is found to be compatible with the ECAL energy cluster position.
The efficiency of the photon identification is measured from $\Z \to \Pe\Pe$ data~\cite{cite:tagandprobe}
by treating the electrons as photons~\cite{long}, and is found to be 90\% for photons with $\pt > 40$\GeV.
These efficiencies include the losses due to
photon conversions caused by the pixel tracker veto requirement,
estimated with $\Z\to\mu\mu\gamma$ events,
where the photon is produced via final-state radiation.

Isolation requirements are based on objects reconstructed with the PF algorithm within
$\Delta R = \sqrt{\smash[b]{(\Delta\eta)^2 + (\Delta\phi)^2}} = 0.3$
from the photon candidate direction, where $\Delta\eta$ and $\Delta\phi$ are,
respectively, the
differences in the pseudorapidity and azimuth angles between the photon and the given reconstructed object.
Only charged candidates are considered in the enforcement of
isolation criteria in the analysis of 13\TeV data,
 whereas additional  photons
are also considered in the analysis of 8\TeV data.

Electron candidates are reconstructed as
clusters of energy deposits in the ECAL matched to signals in the silicon tracker~\cite{CMS:2013ele}.
The electron energy resolution is improved by using a multivariate regression
technique resulting in improvements of 10 and 30\% in the
mass resolution for $\Z\to \Pe\Pe$  events over the standard CMS electron
reconstruction in the barrel and endcap calorimeters, respectively~\cite{CMS:2013ele}.
Electrons are identified via loose requirements
on the shape of these energy deposits, on the ratio of energies in
associated hadron and electromagnetic
calorimeter cells, on the geometrical  matching between the energy
deposits and the associated track,
and on the consistency between the energy reconstructed from the
calorimeter
deposits and the
momentum measured in the tracker.
The electron selection criteria used in the analysis of 8\TeV data are optimized further
 for background rejection
using a multivariate approach.
The training of the multivariate electron reconstruction is performed using simulated events, while the
performance is validated using data.

Muon candidates~\cite{muons} are reconstructed from tracks found in
the muon system that are  associated with the tracks in the silicon detectors.
Muon identification criteria are based on the quality of the track
fit and the number of associated energy deposits
 in the pixel and strip tracking detectors.
The total efficiencies for the combined muon identification
and pileup-corrected isolation criteria are better than 95\%.

Electrons and muons from $\Z$ boson decays are expected to be
isolated from other particles.
A fixed cone of size $\DR  = 0.4$  is
constructed around the direction of each
lepton candidate in the search performed in 8\TeV data,
while $\DR$ varies with the lepton \pt\ in the selection used
in the analysis of 13\TeV data according to the relation:
\begin{linenomath}
\begin{equation}\begin{aligned}
  \Delta R &=\begin{cases}
    0.2, & \pt < 50\GeV\\
    \frac{10\GeV}{\pt}, & 50 < \pt < 200\GeV\\
    0.05, & \pt > 200\GeV.
  \end{cases}\label{eq:miniIso}
\end{aligned}\end{equation}
\end{linenomath}
This ensures high lepton identification efficiency even for highly-boosted $Z$ boson
decays,
as expected in the decay of high-mass resonances.
The relative isolation of the lepton is quantified by summing the transverse momenta of
the relevant PF candidates within this cone, excluding the lepton itself.
To account for the contamination from pileup interactions,
charged particles originating from additional
vertices are excluded from the estimate,
and a correction is applied to  account for
the neutral PF objects originating from
pileup activity, which cannot be excluded by vertex identification.
The resulting quantity, divided by the lepton transverse momentum, is required to be less than
0.4 for both electrons and muons in the analysis of 8\TeV data,
and less than 0.1~(0.2) for electrons~(muons) in 13\TeV data. This requirement
rejects misidentified leptons and  background arising  from hadronic jets.
Finally, the separation between each lepton and the photon must satisfy $\DR> 0.4$
in order to reject events with final-state radiation.

The invariant mass of the dilepton system is required to be greater than 50\GeV.
In the selection used in 8\TeV data, no upper dilepton mass condition is needed,
while in the selection used in 13\TeV data the dilepton mass is required to be below 130\GeV.
The minimum dilepton mass requirement rejects contributions from $\Pp\Pp\to\gamma\gamma^*$, where an internal conversion
of the photon produces a dilepton pair.
In the rare cases where more than one dilepton pair is present, the one with an invariant mass closest to the $\Z$ boson mass is taken.
The final set of requirements combines the information from the
photon and the leptons:
(i)\,the invariant mass $\mzg$ of the $\ell^+\ell^-\gamma$ system~(where $\ell = \Pe,\mu$),
 is required to be above 150\,(200)\GeV in the analysis of 8\,(13)\TeV data; and
(ii)\,the ratio of the photon transverse energy to $\mzg$ must be greater than 0.27.
This latter requirement suppresses backgrounds due to misidentification of photons,
 without significant loss in signal sensitivity and without introducing a bias
in the $\mzg$ spectrum.

\section{Background modelling}

Simulations indicate that 80--90\% of the background after
the full event selection is due to SM $\Z\gamma$ production with initial-state
radiation, with the remainder mostly due to the contribution
from \Z plus jet events, where the jet is misreconstructed as a photon.
The $\mzg$ distributions are steeply and smoothly falling with increasing mass.
The  background is measured directly in the data,
through an unbinned
maximum-likelihood fit to the observed $\mzg$ distributions, separately in the
$\eeg$ and $\mmg$ channels.
The background is parametrized with empirical formulae.

\begin{figure}[tb]
  \centering
    \includegraphics[width=0.45\textwidth]{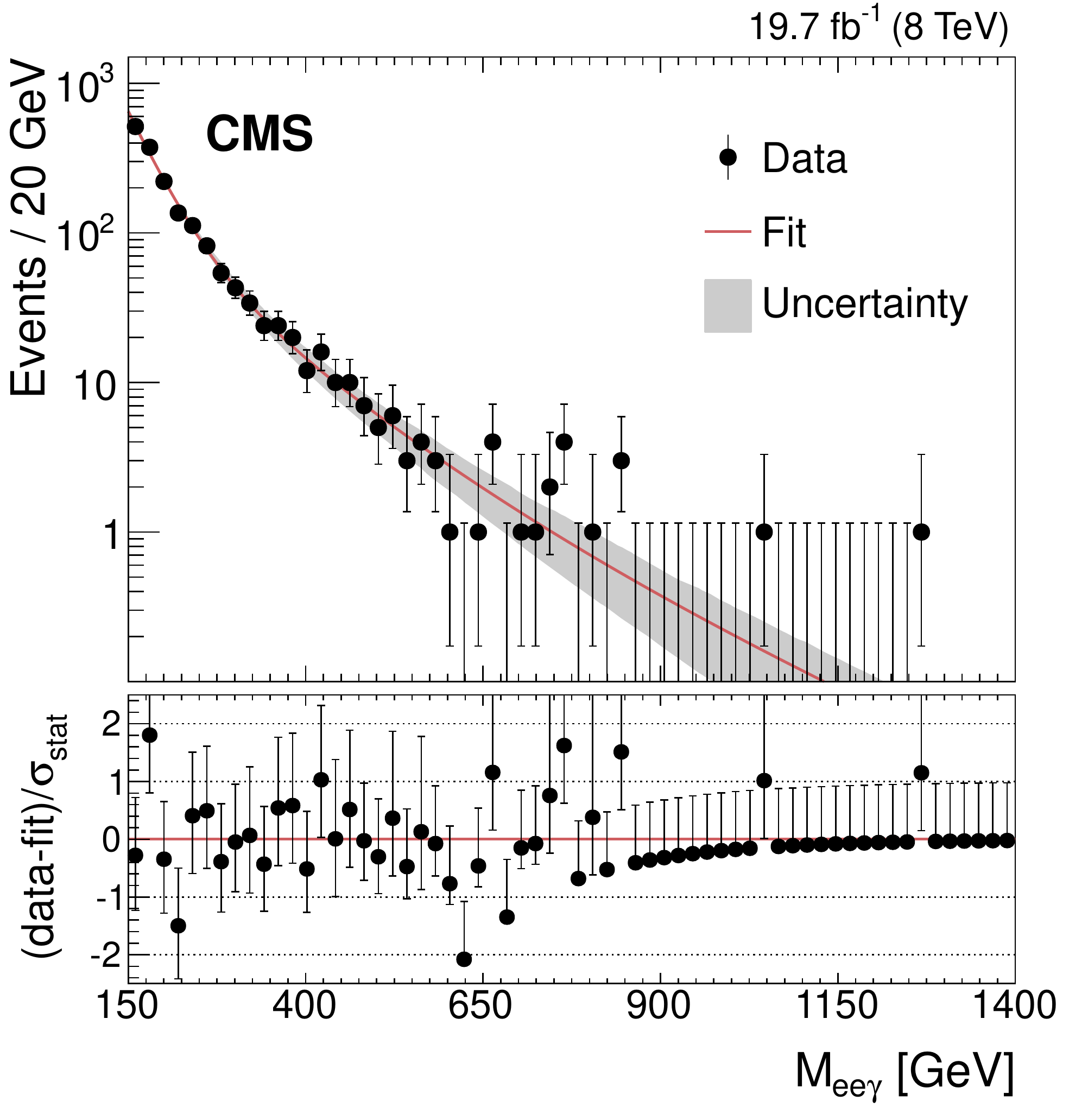}
    \includegraphics[width=0.45\textwidth]{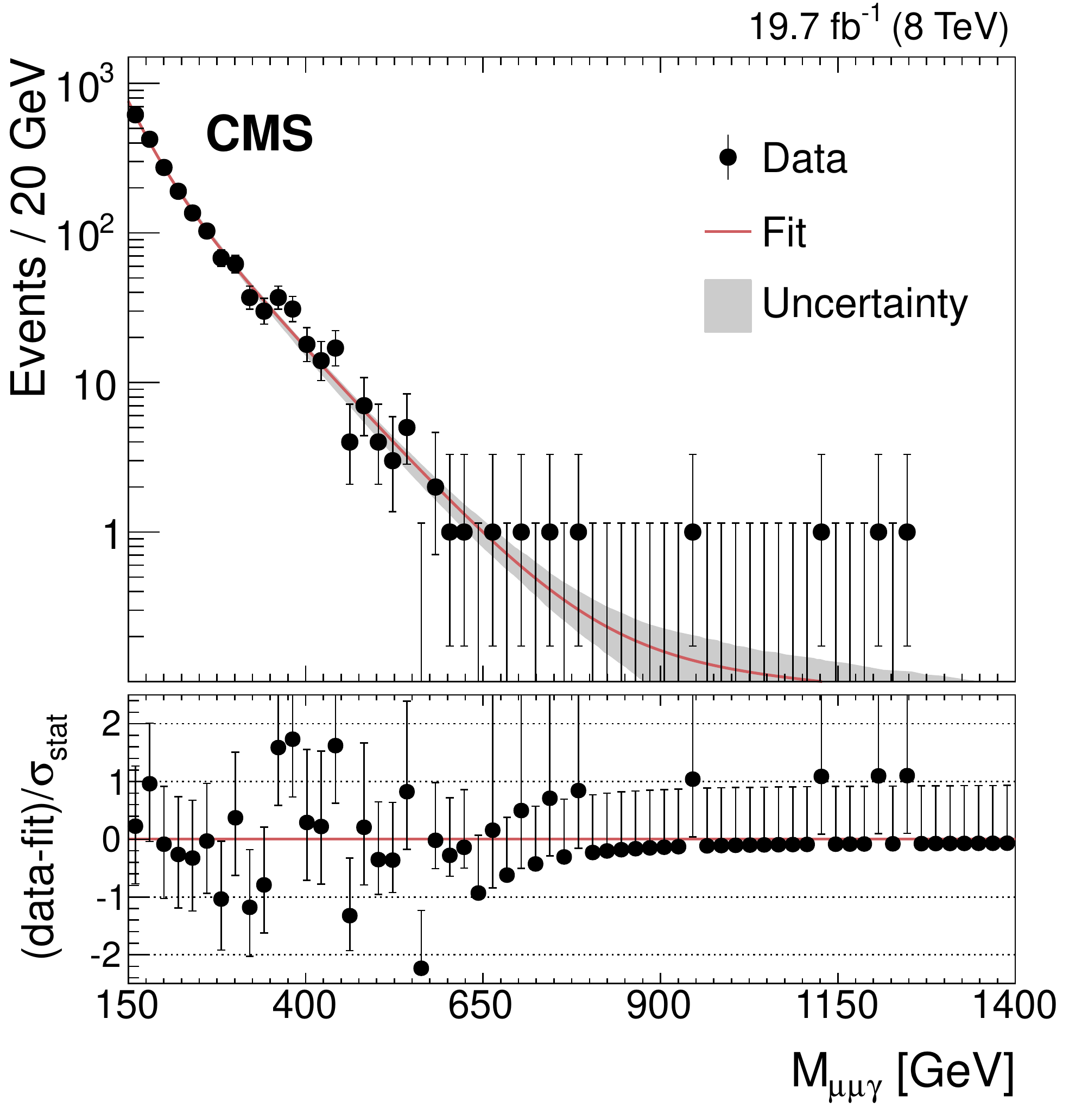}
    \caption{Observed $\mzg$ invariant mass spectra in the 8\TeV data,
for the $\eeg$~(left) and the $\mmg$~(right) channels.
The fitted function is represented by a
line, with the 68\% uncertainty band as grey shading.
The lower panels show the difference between the data and the fit,
divided by the uncertainty $\sigma_\text{stat}$, that includes the statistical uncertainty in both the data and the
fit. For bins with a low number of data entries, the error bars correspond to the Garwood confidence intervals.
    }
    \label{fig:8tev_fits}
\end{figure}

In the 8\TeV analysis the background shape is parameterized with the sum
of three exponential decay functions. The fit is performed for values of
$\mzg> 150$\GeV.
The potential bias in the background measurement is studied by using
pseudo-data generated from different functional forms and fitted with the
function under test.
The results of these fits are used to determine an appropriate
model for the background, such that the bias introduced in the signal
 measurement is smaller than 1/5 of the statistical uncertainty
in its determination.
The chosen model (sum of three
exponential decay functions)
is found to satisfy this criterion across the search mass range.
The observed $\mzg$ invariant mass spectra in 8\TeV data
are shown in Figure~\ref{fig:8tev_fits}.
The results of the fit is represented by a
line, with the 68\% uncertainty band as grey shading.

The 13\TeV search employs
a strategy similar to the 8\TeV search. The fit is performed for values of $\mzg > 200$\GeV.
The function chosen for the background estimate,
\begin{linenomath}
\begin{equation}
f(x) = x^{a + b\log x},
\end{equation}
\end{linenomath}
describes the background shape well
and does not create a significant bias.
The absence of significant bias has been verified
by fitting a large number of pseudo-datasets generated from various
background models, and measuring the difference between the true and fitted
background yields in different $\mzg$ windows; in each window a pull variable is
defined as the difference between the true and fitted yields, divided
by the statistical uncertainty.
If the absolute value of the median of
this distribution is found to be above 0.5 in an interval, an additional uncertainty is assigned to
the background parametrization. A modified pull distribution is then constructed, increasing
the statistical uncertainty in the fit by an extra term, denoted the \textit{bias term}. The bias term is
parametrized as a smooth function of $\mzg$, which is tuned in such a manner that the absolute
value of the median of the modified pull distribution is less than 0.5 in all intervals. This additional
 uncertainty is included in the likelihood function by adding to the background model
a component having the same shape as the signal, with a normalization coefficient distributed
as a Gaussian of mean zero, and with a width equal to the integral of the bias term.
This inclusion of the additional component takes into account the possible
mismodeling of the background shape.
The bias term which is used in this analysis amounts
to about $5\times 10^{-3}$ events/\GeVns{} at $\mzg = 600$\GeV, and smoothly falls to about
$5\times 10^{-4}$ events/GeV around $\mzg=2\TeV$.

\begin{figure}[tb]
  \centering
    \includegraphics[width=0.45\textwidth]{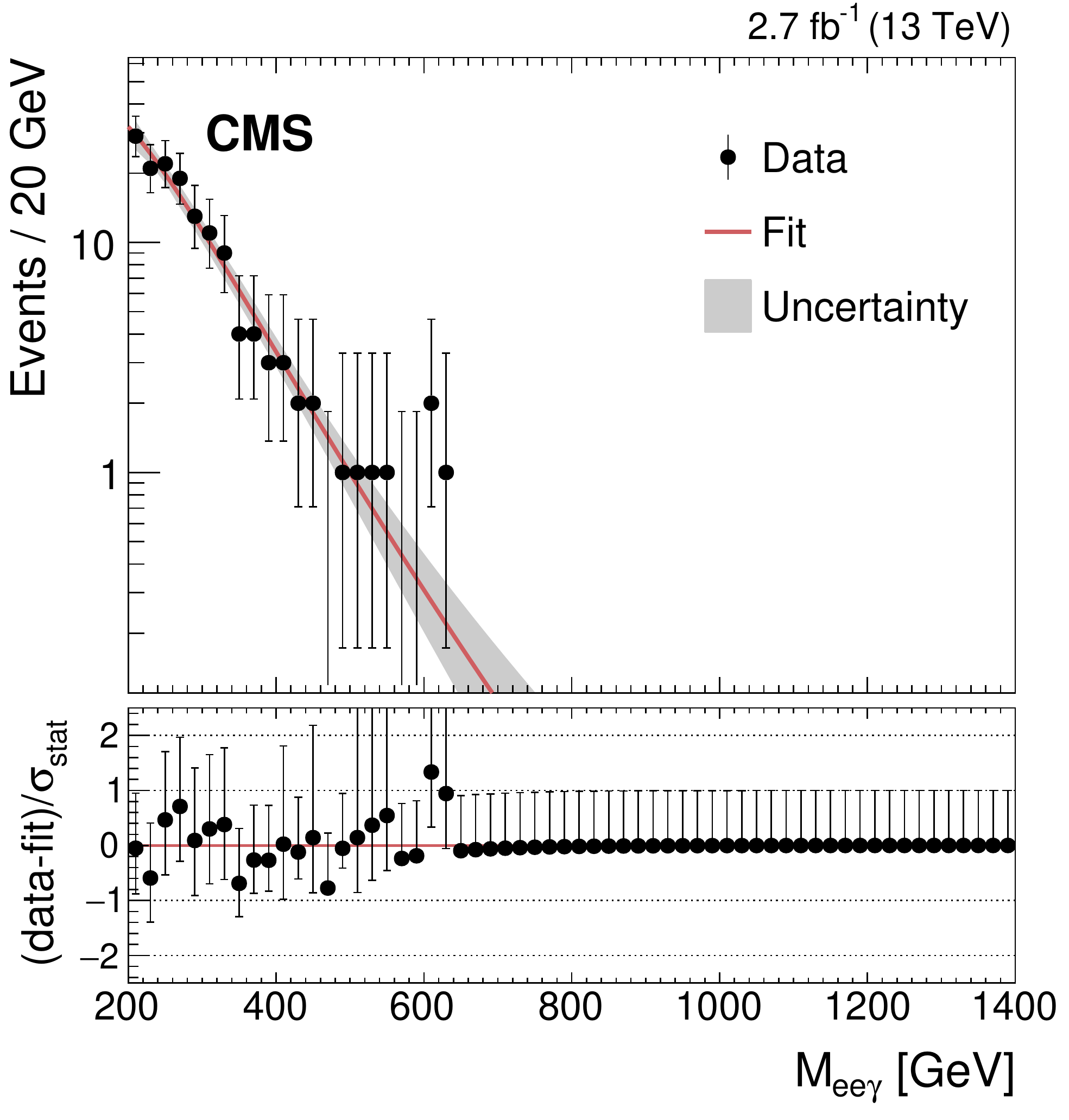}
    \includegraphics[width=0.45\textwidth]{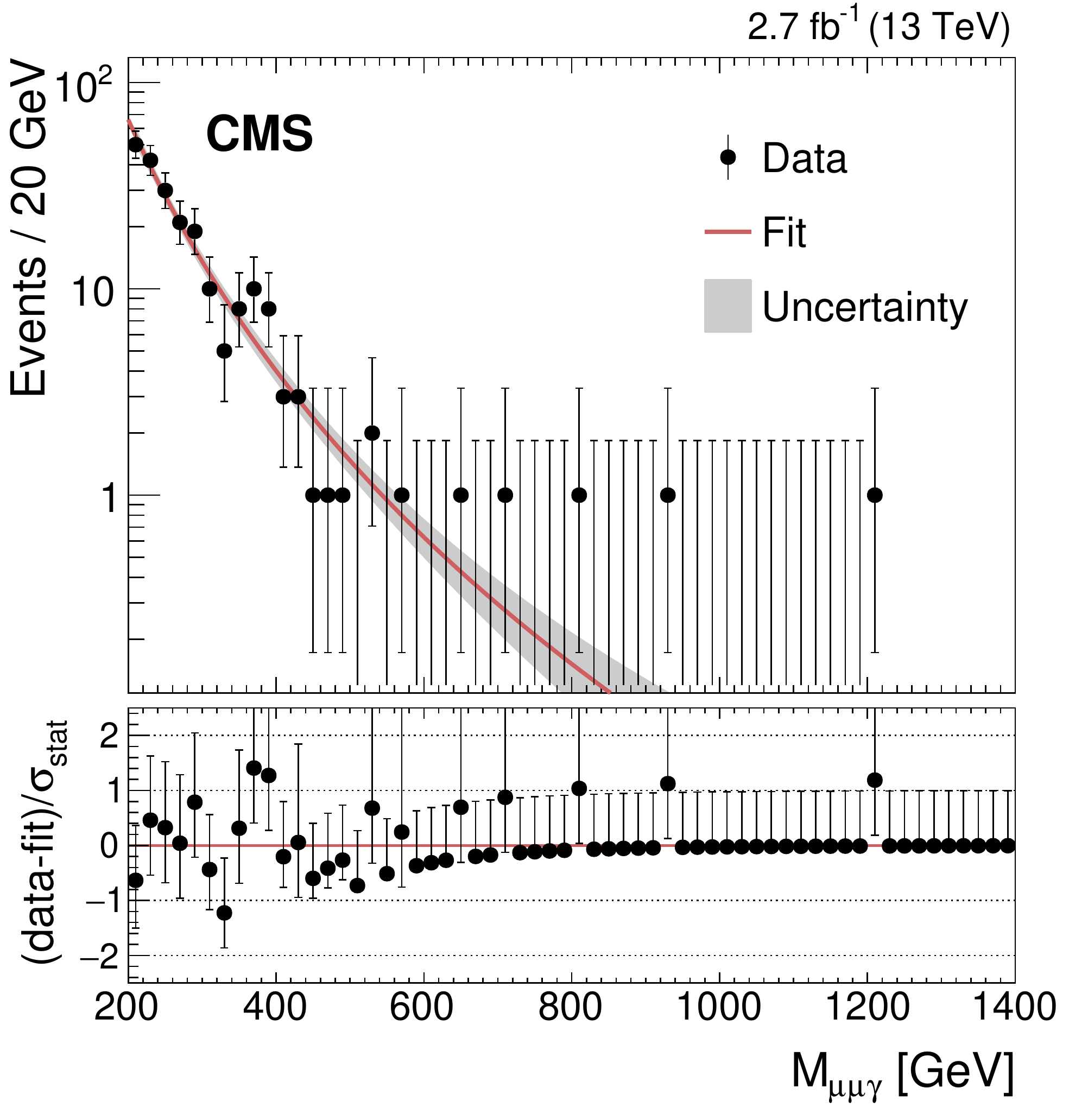}
    \caption{Observed $\mzg$ invariant mass spectra in the 13\TeV data,
for the $\eeg$~(left) and the $\mmg$~(right) channels.
The fitted function is represented by a
line, with the 68\% uncertainty band as gray shading.
The lower panels show the difference between the data and the fit,
divided by the uncertainty $\sigma_\text{stat}$, which includes the statistical uncertainty in both the data and the
fit. For bins with a low number of data entries, the error bars correspond to the Garwood confidence intervals.
    }
    \label{fig:13tev_fits}
\end{figure}

The observed $\mzg$ invariant mass spectra in 13\TeV data are shown in Fig.~\ref{fig:13tev_fits},
for the $\eeg$~(left) and $\mmg$~(right) channels.
The results of the fit and its uncertainty are shown with
a line and a band.

No events with invariant mass larger than 1275\,(1220)\GeV pass the selection
on 8\,(13)\TeV data.

\section{Signal modeling}

We focus on narrow-width signal models, where the intrinsic
width of the resonance is negligible
compared to the experimental resolution.
Scalar resonances decaying to $\Z\gamma$ are generated
at leading order with
\PYTHIA 8.175~\cite{Sjostrand:2007gs}
and  NNPDF2.3~\cite{nnpdf23} parton distribution functions (PDF).
The  8\TeV generator uses the Z2* tune~\cite{z2} to describe the
underlying event and the 13\TeV generator, the CUETP8M tune~\cite{cuetp8m}.
Several samples are generated with masses ranging from 200\,(350)\GeV to
1.2\,(2)\TeV, in the 8\,(13)\TeV analysis.
The search performed in 13\TeV data begins at higher invariant mass
in order to avoid the region where the background is sculpted
by the kinematic selections imposed on the final-state objects.
As far as the upper range, the analysis of the 8\TeV data ends where
the results based on the 13\TeV analysis dominate the combination.

The signal distribution in $\mzg$ is obtained from the generated
events  that pass the full selection.
The signal shape is parametrized with empirical functions;
the function chosen is the sum of a Gaussian and
Crystal Ball function~\cite{crystalball} for the 8\TeV analysis,
and an extended form of the Crystal Ball function, with a Gaussian
core and two power-law tails, for the 13 TeV analysis.
The fitted parameters are determined from the simulated
 samples at each mass point, separately for the electron
and muon channels,
and then interpolated through polynomial fits to generic
$\mzg$ values in order to
have smoothly varying signal shape parametrizations.
The typical mass resolution for signal events is 1\% for the $\eeg$ channel
and 1--2\% for the $\mmg$ channel, depending on the mass of the resonance.

The product of the expected signal acceptance and efficiency
in the analysis of 8\TeV data rises from about 33\% at
$\mzg = 200$\GeV to about 45\% at $\mzg = 1.2$\TeV.
In the analysis of 13\TeV data it
rises from about 25\%~(35\%) at
$\mzg = 350$\GeV to about 45\%~(55\%) at $\mzg = 2$\TeV, for
the $\eeg$~($\mmg$) channel.

\section{Systematic uncertainties}

The background spectra are described by parametric functions of $\mzg$. The coefficients are
obtained from a fit to the data events, and considered as unconstrained nuisance parameters in
the fit. Thus the description of the background is derived from data.
No systematic uncertainty related to the background description is considered,
as possible biases are accounted for in the bias terms.

The systematic uncertainty in the signal description
arises from the integrated  luminosity
measurement~\cite{CMS-PAS-LUM-13-001,lumi}, the trigger efficiency,
the effect on the signal acceptance from the choice
 of parton distribution functions~\cite{pdf4lhc},
the imperfect simulation of the lepton and photon efficiencies,
and the signal mass scale and resolution. These uncertainties have been
evaluated separately at 8 and 13\TeV, and their magnitudes are summarized
in Table~\ref{tab:syst}. The photon efficiency uncertainty of the 13\TeV
data analysis is larger because of the use of preliminary calibrations.
The sources of uncertainty are considered to be
completely uncorrelated between the two center-of-mass energies.

\begin{table}[htbp]
\topcaption{Summary of considered systematic uncertainties in signal. \label{tab:syst}}
\centering
\begin{tabular}{lcc}
\hline
Source     & 8\TeV & 13\TeV \\
\hline
Integrated luminosity & 2.6\% & 2.7\%\\
PDF choice & 1\% & 1\%\\
Trigger efficiency (ee, $\mu\mu$) & 3\%, 2\% & 3\%, 2\% \\
Lepton efficiency & 5\% & 5\% \\
Photon efficiency & 1--2.6\% & 5\% \\
Mass scale and resolution (ee$\gamma$, $\mu\mu\gamma$) & 1\%, 1--10\% & 1\%, 1--5\% \\
\hline
Total systematic uncertainty (ee$\gamma$, $\mu\mu\gamma$) & 6.6--7.0\%, 6.2--12\% & 8.3\%, 8.3--9.6\% \\
\hline
\end{tabular}
\end{table}

\section{Results}

\begin{figure}[hbt]
  \centering
   \includegraphics[width=1.2\cmsFigWidth]{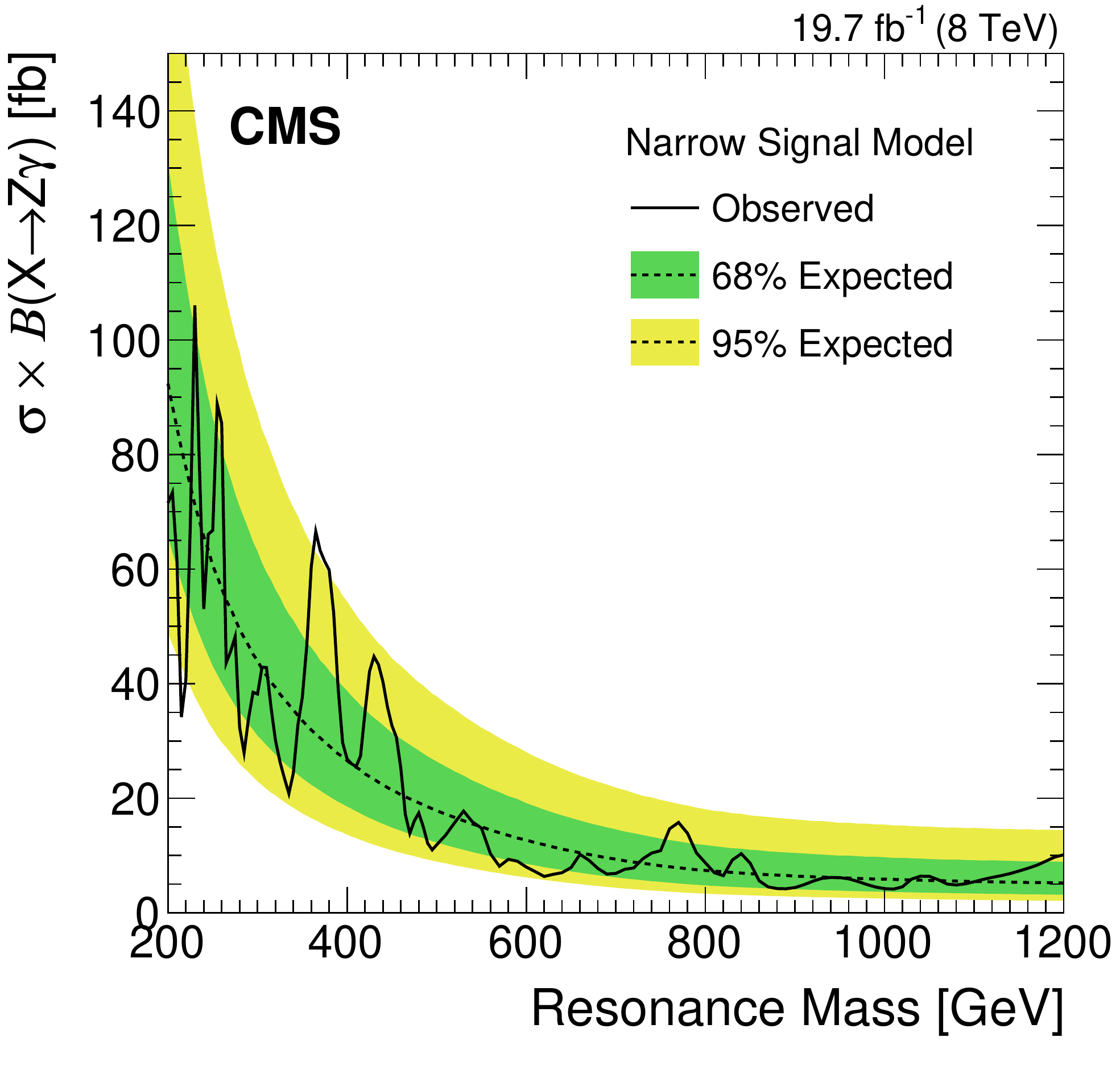}
   \includegraphics[width=1.2\cmsFigWidth]{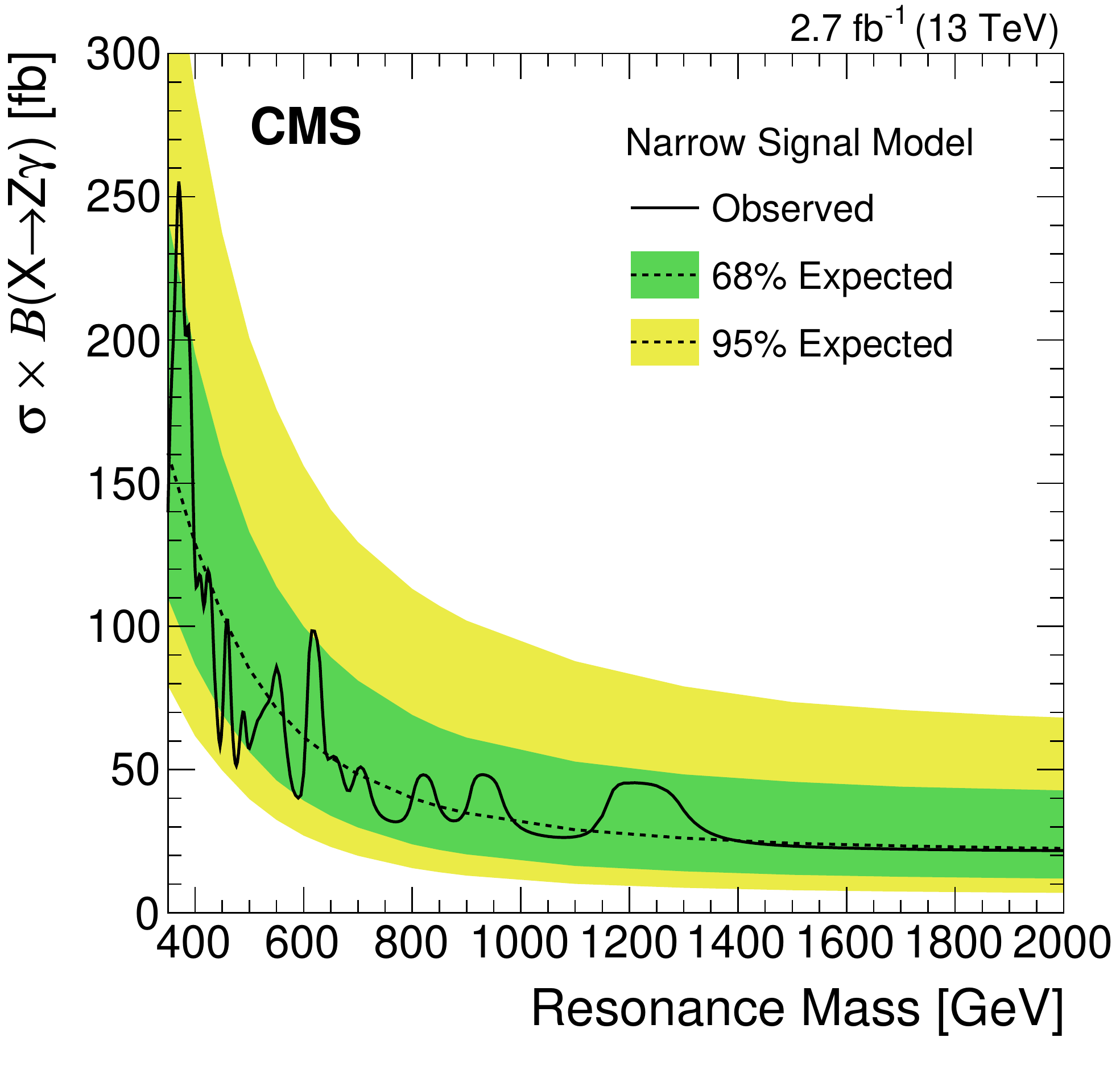}
    \caption{Expected and observed upper limits, at 95\% CL, on
the cross section times branching fraction for
${\PX}\to\Z\gamma$
obtained with the searches
performed at 8\TeV~(left) and at 13\TeV~(right).
}
    \label{fig:rawlimit}
\end{figure}

No significant excess is observed with
respect to the SM background predictions. Upper limits are set
on the production cross section of high-mass
scalar resonances
using the modified frequentist method, commonly known as CL$_\mathrm{s}$~\cite{cls2,Junk:1999kv}.
An example of its usage is found in \cite{cmsHiggs}.
Asymptotic formulae \cite{asympt} are
used in the calculation.
The individual expected and observed upper limits
at 95\% confidence level~(CL) on the product of the cross section and the
branching fraction for ${\PX}\to\Z\gamma$
are shown in Fig.~\ref{fig:rawlimit}.

The combination of the two results accounts for  the different
parton luminosities for collisions at 8 and 13\TeV,
which have been calculated with the NNPDF2.3 parton distributions~\cite{nnpdf23}.
The effect of using different PDFs for the scaling has been evaluated
and affects the limits by  at most a  few percent, mainly in the low-mass region.
The signal is assumed
to be produced solely through gluon-gluon fusion, and the 8\TeV limit is
scaled up by the corresponding parton luminosity ratio,
which ranges between 3 and 7 in the 0.2 to 1.2\TeV mass region, and is about
4.3 for a signal with a mass of 750\GeV.

\begin{figure}[tb]
  \centering
    \includegraphics[width=1.2\cmsFigWidth]{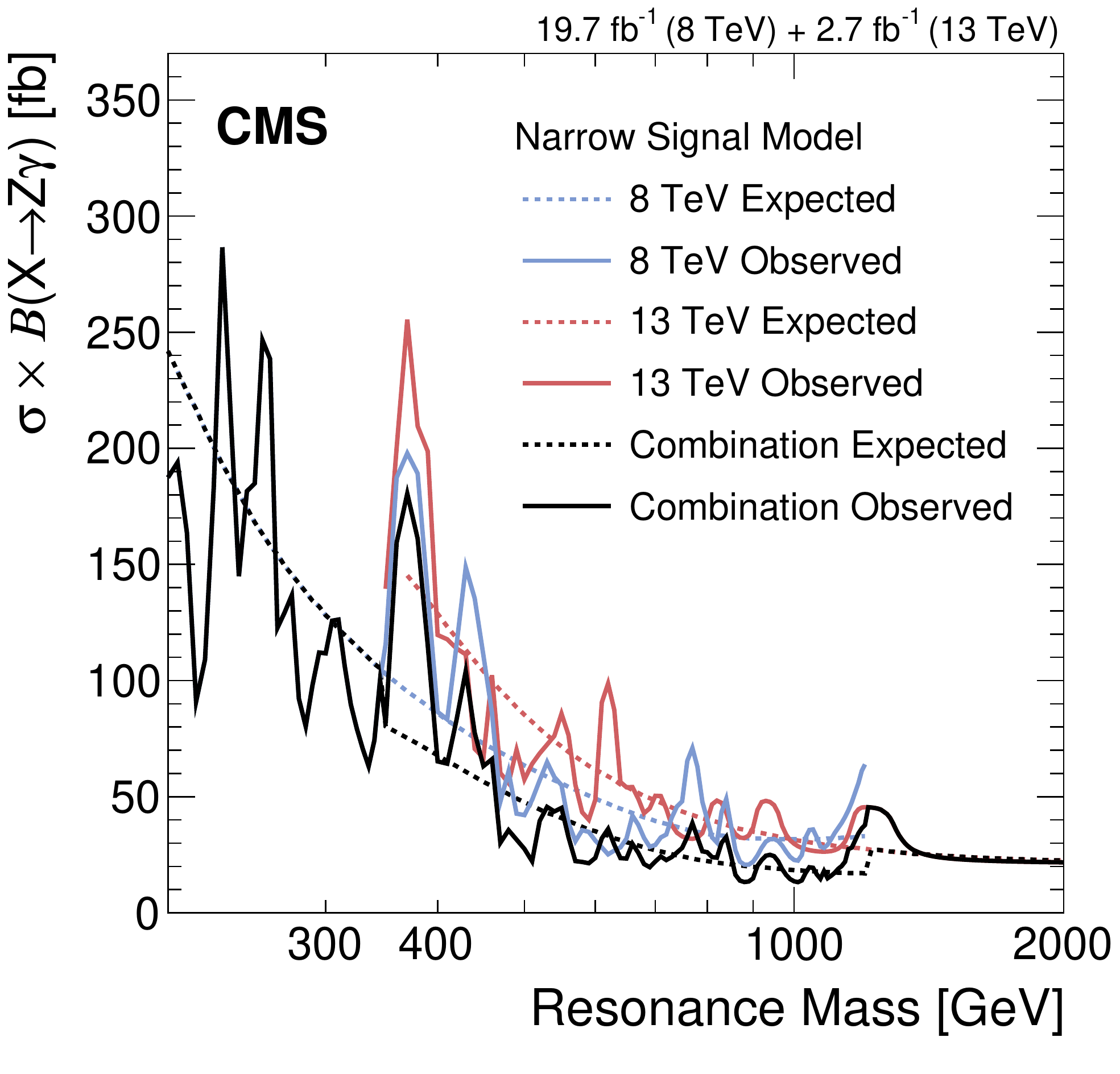}
    \includegraphics[width=1.2\cmsFigWidth]{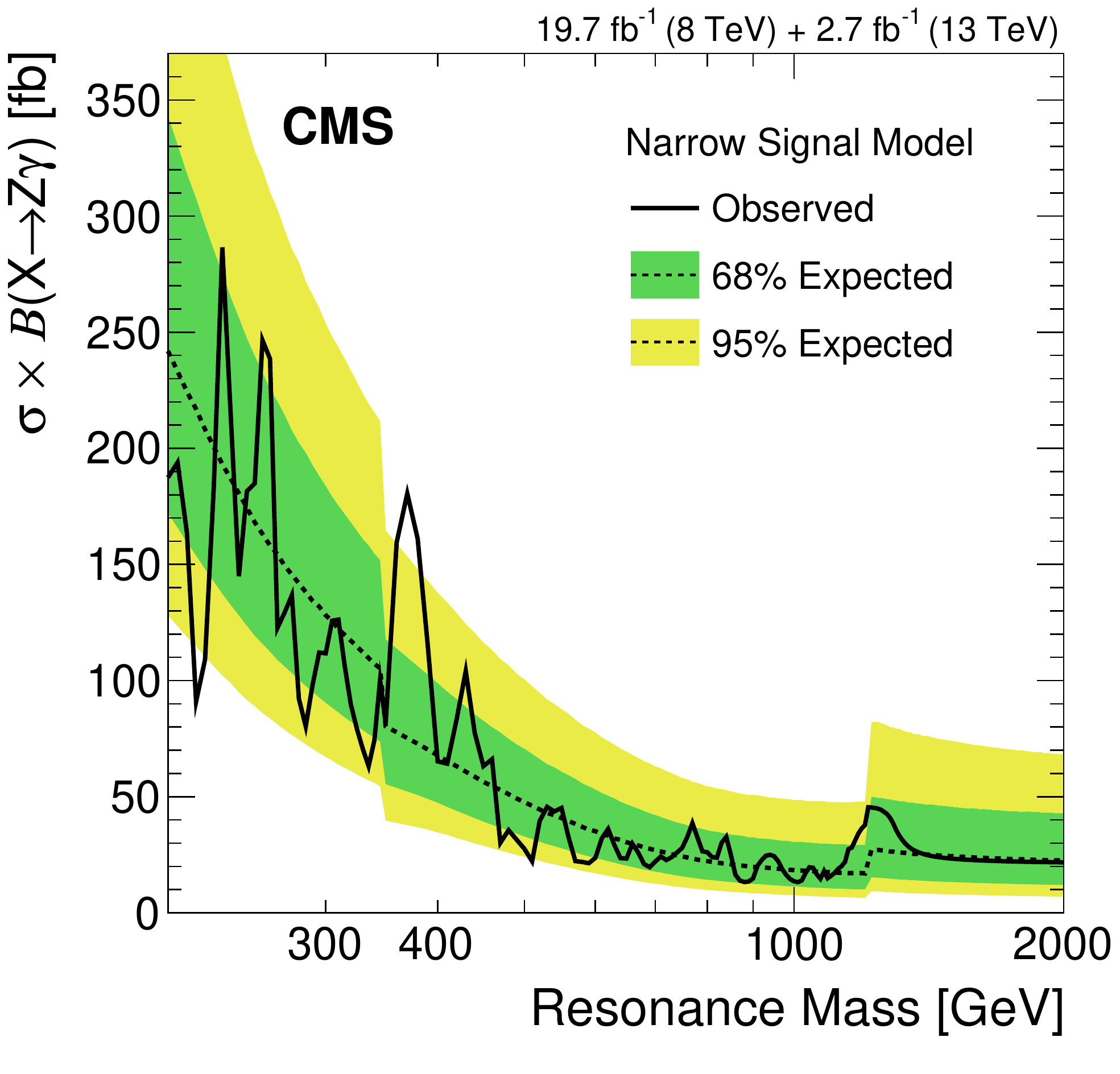}
    \caption{Left: expected and observed upper limits, at 95\% CL,
on the 13\TeV cross section
$\sigma_{13\TeV}({\PX} \to\Z\gamma)$
for the scaled 8\TeV~(blue, lighter) and 13\TeV~(red, darker) searches,
together with their combination~(black).
Expected limits are shown with dashed lines,
observed ones with solid lines.
Right:
95\% CL upper limit for the combination of 8\TeV and 13\TeV data.
The solid~(dashed) line represents the observed~(expected) limit,
whereas the inner green~(outer yellow) bands represent the 68\%~(95\%) uncertainty
bands.
    }
    \label{fig:ul}
\end{figure}

\begin{figure}[tb]
  \centering
    \includegraphics[width=0.45\textwidth]{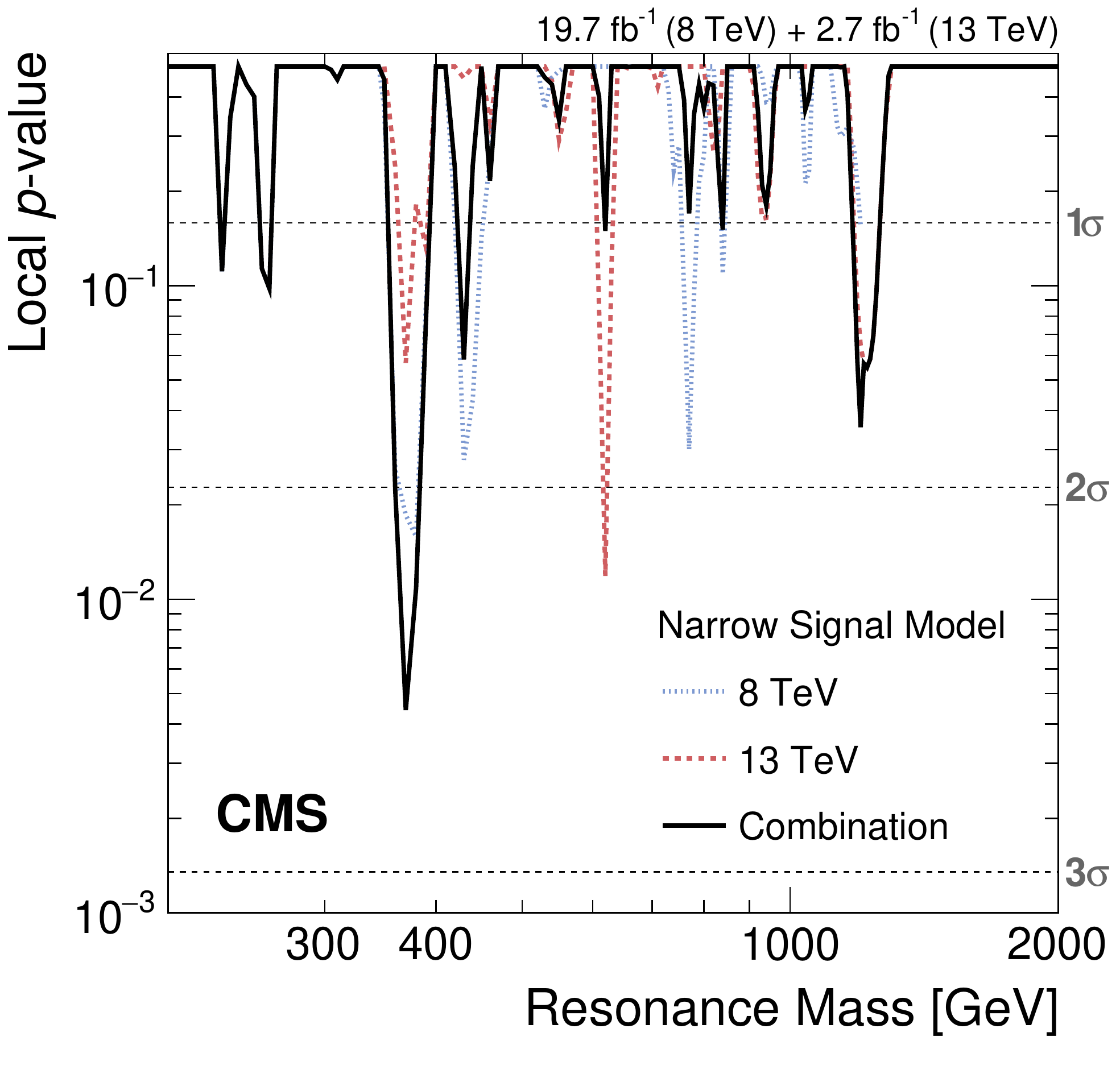}
    \caption{Observed background-only local $p$-values for the scaled
8\TeV search~(blue, dotted), the
13\TeV search~(red, dashed), and the combination~(black, solid).
    }
    \label{fig:pvalue}
\end{figure}

Figure~\ref{fig:ul}~(left) shows the 95\% CL upper limits on
the  13\TeV cross section,
$\sigma_{13\TeV}({\PX} \to\Z\gamma)$, as a function
of the resonance mass, for the 8\TeV~(blue, lighter) and 13\TeV~(red, darker) analyses, and their combination~(black).
The expected (observed) limits are shown as dashed (solid)  lines.
Figure~\ref{fig:ul}~(right)  shows the combined 8 and 13\TeV limit with its 68\%~(inner green) and
95\%~(outer yellow) uncertainty bands. The discontinuities in the limits are an artifact
of the different ranges exploited by the two searches.

Background-only local $p$-values are defined as the probability of obtaining, under the
back\-ground-only hypothesis, a result equal or larger than the one observed in the data.
Figure~\ref{fig:pvalue} shows the observed background-only $p$-values for the 8\TeV search~(blue, dotted), the
13\TeV search~(red, dashed), and their combination~(black). The fluctuation at $\mzg \approx 370$\GeV
corresponds to a local significance of 2.6~$\sigma$, and a global significance smaller than one
standard deviation,
once the `look-elsewhere' effect has been taken into account~\cite{lee}.
This has been computed by counting the fraction of times the background-only $p$-value crosses
the level corresponding to 0.5 standard deviations in the full mass range in which limits are set.

\section{Summary}
\label{sec:conclusions}

A search for heavy resonances decaying to $\Z\gamma$,
with further decay $\Z\to \ell^+\ell^-$, with $\ell = \Pe$ or
$\mu$, has been presented.
The search makes use of proton-proton data collected by the CMS
detector at the LHC, corresponding to integrated
luminosities of 19.7 and 2.7\fbinv at 8 and 13\TeV, respectively.
The background is measured directly from data and localized excesses are looked for.
No significant deviation with respect to the standard model expectation is found.
Upper limits at 95\% confidence level are set on the production cross section of
narrow resonances,
ranging from 280 to 20\unit{fb} for resonance masses from
200 to 2000\GeV.

\begin{acknowledgments}
We congratulate our colleagues in the CERN accelerator departments for the excellent performance of the LHC and thank the technical and administrative staffs at CERN and at other CMS institutes for their contributions to the success of the CMS effort. In addition, we gratefully acknowledge the computing centers and personnel of the Worldwide LHC Computing Grid for delivering so effectively the computing infrastructure essential to our analyses. Finally, we acknowledge the enduring support for the construction and operation of the LHC and the CMS detector provided by the following funding agencies: BMWFW and FWF (Austria); FNRS and FWO (Belgium); CNPq, CAPES, FAPERJ, and FAPESP (Brazil); MES (Bulgaria); CERN; CAS, MoST, and NSFC (China); COLCIENCIAS (Colombia); MSES and CSF (Croatia); RPF (Cyprus); SENESCYT (Ecuador); MoER, ERC IUT and ERDF (Estonia); Academy of Finland, MEC, and HIP (Finland); CEA and CNRS/IN2P3 (France); BMBF, DFG, and HGF (Germany); GSRT (Greece); OTKA and NIH (Hungary); DAE and DST (India); IPM (Iran); SFI (Ireland); INFN (Italy); MSIP and NRF (Republic of Korea); LAS (Lithuania); MOE and UM (Malaysia); BUAP, CINVESTAV, CONACYT, LNS, SEP, and UASLP-FAI (Mexico); MBIE (New Zealand); PAEC (Pakistan); MSHE and NSC (Poland); FCT (Portugal); JINR (Dubna); MON, RosAtom, RAS and RFBR (Russia); MESTD (Serbia); SEIDI and CPAN (Spain); Swiss Funding Agencies (Switzerland); MST (Taipei); ThEPCenter, IPST, STAR and NSTDA (Thailand); TUBITAK and TAEK (Turkey); NASU and SFFR (Ukraine); STFC (United Kingdom); DOE and NSF (USA).

\hyphenation{Rachada-pisek} Individuals have received support from the Marie-Curie program and the European Research Council and EPLANET (European Union); the Leventis Foundation; the A. P. Sloan Foundation; the Alexander von Humboldt Foundation; the Belgian Federal Science Policy Office; the Fonds pour la Formation \`a la Recherche dans l'Industrie et dans l'Agriculture (FRIA-Belgium); the Agentschap voor Innovatie door Wetenschap en Technologie (IWT-Belgium); the Ministry of Education, Youth and Sports (MEYS) of the Czech Republic; the Council of Science and Industrial Research, India; the HOMING PLUS program of the Foundation for Polish Science, cofinanced from European Union, Regional Development Fund, the Mobility Plus program of the Ministry of Science and Higher Education, the National Science Center (Poland), contracts Harmonia 2014/14/M/ST2/00428, Opus 2013/11/B/ST2/04202, 2014/13/B/ST2/02543 and 2014/15/B/ST2/03998, Sonata-bis 2012/07/E/ST2/01406; the Thalis and Aristeia programs cofinanced by EU-ESF and the Greek NSRF; the National Priorities Research Program by Qatar National Research Fund; the Programa Clar\'in-COFUND del Principado de Asturias; the Rachadapisek Sompot Fund for Postdoctoral Fellowship, Chulalongkorn University and the Chulalongkorn Academic into Its 2nd Century Project Advancement Project (Thailand); and the Welch Foundation, contract C-1845. \end{acknowledgments}
\bibliography{auto_generated}

\cleardoublepage \appendix\section{The CMS Collaboration \label{app:collab}}\begin{sloppypar}\hyphenpenalty=5000\widowpenalty=500\clubpenalty=5000\input{EXO-16-021-authorlist.tex}\end{sloppypar}
\end{document}

%% file: EXO-16-021-authorlist.tex
\textbf{Yerevan Physics Institute,  Yerevan,  Armenia}\\*[0pt]
V.~Khachatryan, A.M.~Sirunyan, A.~Tumasyan
\vskip\cmsinstskip
\textbf{Institut f\"{u}r Hochenergiephysik,  Wien,  Austria}\\*[0pt]
W.~Adam, E.~Asilar, T.~Bergauer, J.~Brandstetter, E.~Brondolin, M.~Dragicevic, J.~Er\"{o}, M.~Flechl, M.~Friedl, R.~Fr\"{u}hwirth\cmsAuthorMark{1}, V.M.~Ghete, C.~Hartl, N.~H\"{o}rmann, J.~Hrubec, M.~Jeitler\cmsAuthorMark{1}, A.~K\"{o}nig, I.~Kr\"{a}tschmer, D.~Liko, T.~Matsushita, I.~Mikulec, D.~Rabady, N.~Rad, B.~Rahbaran, H.~Rohringer, J.~Schieck\cmsAuthorMark{1}, J.~Strauss, W.~Waltenberger, C.-E.~Wulz\cmsAuthorMark{1}
% ...
\vskip\cmsinstskip
% ...
1:~~Also at Vienna University of Technology, Vienna, Austria\\
% ...
